\documentclass[conference, a4paper]{IEEEtran}
\usepackage{graphicx} % \usepackage[pdftex]{graphicx}
\usepackage{amsmath}
\usepackage{subcaption}
\usepackage[font=footnotesize]{caption}
\usepackage{flushend}

\usepackage[utf8]{inputenc}
\usepackage[T1]{fontenc}
\usepackage{graphicx}
\usepackage{amsfonts,amsmath}

\usepackage{subcaption}
\usepackage{graphicx}
\usepackage{epstopdf}
\usepackage{amsmath}
\usepackage{xcolor}

\usepackage[american]{babel}

\definecolor{review}{RGB}{0,0,0}   % rot

\definecolor{review_2}{RGB}{0,0,255}  % blau
\definecolor{review_3}{RGB}{255,0,0}  % red

\colorlet{bvb}{black}
\colorlet{wz}{black}

\def\BibTeX{{\rm B\kern-.05em{\sc i\kern-.025em b}\kern-.08em T\kern-.1667em\lower.7ex\hbox{E}\kern-.125emX}}

\begin{document}

\title {Start Stop Bit Method for Efficient Data Communication in 6G Mobile Radio Systems}

\author{Wolfgang Zirwas, Berthold Panzner, Rakash Sivasivaganesan, Brenda Vilas Boas, Luis A. Su\'arez \\
\IEEEauthorblockA{Nokia Standards, Munich, Germany \\
		Email: wolfgang.zirwas@nokia.com}}

\maketitle

\begin{abstract}

In this article, a novel approach for mobile radio communications is proposed and analysed, which is promising for future $6$G cooperative distributed MIMO systems. The fundamental idea is a new mechanism namely `\emph{start stop} bit' method, which transmits bit sequences as the \emph{start/stop}  bits of a synchronized counter instead of transmitting the full encoded bit sequence itself. 
 In that way, theoretically, we can transmit infinitely long data messages with only one bit for starting and one bit for stopping the counter. The value of the counter, as identified by the stop bit, is then used to reconstruct and remap the one and unique transmitted bit sequence. 
 The start stop bit method is characterized by a high signal sparsity as only two bits are transmitted, independently of the bit sequence length for the message. Among the benefits of the start stop bit method are energy efficient data transmission, and effective distributed MIMO systems, which exploit the sparse inter cooperation area interference as well as the low processing complexity for the sparse precoder calculation. 
Moreover, for the next mobile wireless generation, we propose an advanced scheme of the start stop bit method which enhances  its resource usage.
        
We call the resulting method a sparse dMIMO system. 

\end{abstract}

\IEEEpeerreviewmaketitle

\section{Introduction}

Future mobile radio systems like $5$G Advanced, or, more importantly $6$G are expected to significantly improve key performance indicators like the supported data rates, the number of served users as well as the spectral efficiency. 
The probably most important challenge is then sustainability and energy efficiency. One relevant part of the overall energy consumption is the transmit signal power $ P_{\mathrm{t,bit}} $ in joule per bit, which has to decrease correspondingly to the constantly increasing data rates.

In parallel, we expect a trend to an increased number of sensor devices, potentially powered by energy harvesting, or, with a single battery of limited capacity. These ultra low power devices need, or, benefit from new methods for data transmission and one discussed option is to use passive devices with varying reflector antennas \cite{TR22840}. Another application benefiting from efficient data communication might be enhanced uplink coverage, where deep coverage is often achieved by relative inefficient excessive repetition coding of transmit data \cite{GSMAambient}. 

In this paper, we focus on distributed MIMO (dMIMO) for $6$G as introduced in \cite{HexaXdMIMO}. $6$G dMIMO systems are expected to have similarities with $5$G joint transmission cooperative multipoint (JT CoMP) and promise higher data rates, coverage gains as well as an improved energy efficiency. Ideally, dMIMO energy efficiency gains can be achieved by mitigation of inter cell interference, the rank enhancement effect, and the inherent beamforming gains of suitably precoded large number of mMIMO antenna elements from multiple transmitter sites. Interference mitigation, rank enhancement, and beamforming gains help to increase the signal to interference and noise ratio (SINR) per data stream. 
Especially, beamforming focuses the transmit power at the intended receiver device instead of spreading it over larger spatial areas. 
Despite these promising benefits of dMIMO, there has been only small to moderate gains reported for JT CoMP system level simulations~\cite{3GPPCoMP}. The main reason is the difficulty to overcome inter cooperation area interference.   

In this paper, we propose the start stop bit method which can significantly lower the inter cooperation area interference in dMIMO system. Moreover, the start stop bit method has a very sparse precoding matrix, where about $98\%$ of the resource elements are zero.

This sparsity can be exploited to lower the processing complexity as well as the related power consumption at the JT CoMP precoder, thereby addressing two main obstacles of current dMIMO systems. 

The reminder of this paper is organized as follows, Section \ref{secII} provides a basic background discussion.  Section \ref{secIII} introduces the basic \emph{start stop} concept and extensions, and Section \ref{secIV}  describes in details the proposed enhancements. In the sequel, Section \ref{secV} describes the idea of a sparse dMIMO system. Finally, Section \ref{sec:conclusion} concludes this paper.

\section{ Background}
\label{secII}

From a system level energy efficiency point of view, there is a significant issue, i.e., the typically high processing complexity for the JT CoMP precoder calculations. 
It involves multiple matrix inversions of the channel matrices $ \mathbf{H}_{i}~\in~C^{N_{K} \times N_{\mathrm{tx}}} $, with $ N_{K} $ as the number of UE receive antennas and $ N_{\mathrm{tx}} $ as the number of transmit antennas of the cooperation area.  
This matrix inversion is often a Moore Penrose matrix inversion $ \mathbf{W}_{i}~=~\mathrm{pinv}(\mathbf{H}_{i}) $, $ \mathbf{W}_{i}~\in~C^{ N_{\mathrm{tx}} \times  N_{K}} $ and has to be calculated for each physical resource block (PRB) $ \mathrm{PRB}_{i} $ or subband $ \mathrm{B}_{i} $. There are $ i~=~1 \ldots I $ PRBs and for a 20~MHz 3GPP new radio (NR) system there are overall $ I~=~100 $ PRBs. 
Such a system has $ j~=1 \ldots J $ subcarriers, where for a 20~MHz system $ J~=~1200 $. For a fully loaded system, each subcarrier $ j $  requires a matrix multiplication  $ \mathbf{W}_{i} \: \mathbf{x}_{j} $ with $ \mathbf{x}_{j}~\in~C^{ N_{K} \times 1} $ for each OFDM symbol, which leads in the case of NR, with $14$ OFDM symbols per transmission time interval (TTI) of $1$~ms, to overall $14000$ matrix multiplications per second and per subcarrier. This overall high JT CoMP processing complexity results in a significant power consumption, which might outweigh any potential power savings due to a more efficient data transmission over the air.

The basic idea of the \emph{start stop} bit method is very simple, i.e., it replaces the transmission of bit sequences by start and stop bits controlling a counter located at the receiver. 
For that purpose, the gNB in downlink, or, the user in uplink sends a first bit as the start bit and then transmits a stop bit when the counter value represents the intended bit sequence. 
That way, each possible bit sequence of length $ N_{\mathrm{b}} $ is associated with one single counter value $ n_{\mathrm{c}}~\le~N_{\mathrm{c}} $. The receiver, for example, a gNB, starts its synchronized counter when receiving the \emph{start bit} and stops it with the \emph{stop bit}. The final counter value can then be remapped to the related transmitted bit sequence. This simple method allows in principle to transmit bit sequences of any length with high power efficiency and occupies the wireless radio channel just for two bits. 

For very short messages of sensor devices with bit sequences, e.g., of length $ N_{\mathrm{b}}~=~10 $ and respective counter length of $  N_{\mathrm{c}}~=~2^{N}~=1024 $ one might apply the proposed scheme directly. Nonetheless, the counter length increases exponentially with the length of the transmitted bit sequence. Hence, implementation aspects like latency will limit the number of bits to  $ N_{\mathrm{b}}~=~10 $ to maybe $ N_{\mathrm{b}}~=~20 $ and the related counter length from $ N_{\mathrm{c}}~=~1024 $ to  $ N_{\mathrm{c}}~>~10^6 $.  

Therefore, as a first extension we need to introduce a scheme were $ M $ messages with bit sequences $ \mathbf{b}_{m} $ of limited length $ N_{\mathrm{b}} $ are transmitted in parallel over one single counter. Then, we need just one single start bit for all $ M $ bit sequences and one additional stop bit per bit sequence $ \mathbf{b}_{m} $, with $ m~=~1~\ldots~M $.

Next, we apply and adapt the proposed scheme to OFDM systems like 3GPP long term evolution (LTE) or similarly 3GPP new radio (NR). For that purpose, we map the stop bit(s) to the resource elements (RE) of the physical resource blocks (PRB) of the OFDM symbols, which has the additional benefit that we might avoid the transmission of any start bit by defining the first resource element of the first PRB as the counter start. 

In a further enhancement, we extend the counter space from the frequency and time domain into the modulation and coding space. For instance, one could assume the transmission of QAM symbols per active resource element, i.e.,  per stop bit. This extension of the counter space is introduced with the goal to increase the data rate efficiency $ \gamma $ in bits per channel use (bpcu), which is defined as the overall number of transmitted bits $ N_{\mathrm{b,sum}} $ divided by the number of resource elements $ R $ needed to transmit these bits. Therefore, these methods intend to increase the user throughput and/or the capacity of the system. But, for the above described means the data rate efficiency $ \gamma_{\mathrm{s}} $ of our scheme is still significantly below that of a conventional OFDM system  $ \gamma_{\mathrm{OFDM}} $, i.e., worse by about a factor of ten. For that reason, we evaluated directions how to increase the counter space further, which resulted in the idea of time- and frequency-shifting of the stop bits. With time shifting we mean the shifting of the transmit symbol timing by $ N_{\mathrm{t}} $ time delays $ \delta_{t} $ relative to the OFDM symbol timing, where  $ \delta_{t} $ is small relative to the overall OFDM symbol length $ T $. Similarly, with frequency shifting, we mean the shifting of the transmit frequency by $ N_{\mathrm{f}} $ frequency offsets $ \delta_{f} $ relative to the correct subcarrier frequency $ f_{\mathrm{SC}} $. Furthermore, these new dimensions can be leveraged when the stop bits are isolated and the related inter symbol and inter carrier interference does not harm any other receive bits. Hence, this is a unique advantage of the \emph{start stop} method.    

Adapted coding methods look promising for the proposed method and will be discussed with respect to achievable coverage gains. 
Finally, we elaborate the beneficial system level aspects like reduced inter cell interference and reduced processing complexity for dMIMO systems, which is the main focus in this paper due to its potential impact on future $6$G systems. 
The \emph{start stop} method might become a main enabler for practical JT CoMP implementations, even though this needs further verification and research. 

We should note that the basic start stop method has similarities to well known techniques such as pulse position modulation (PPM) \cite{PPM}, ultra wideband modulation (UWB) \cite{UWB}, or, subcarrier index modulation (SC-IM) \cite{IM}, which has been inspired by spatial modulation (SM). PPM is a very old and simple analogue technique, where the position of a digital pulse is shifted in time to transfer a certain information like the remote control signals for a small airplane, which is then different to the communication of bit sequences. 

UWB uses ultra narrow time signal pulses generated from a very large bandwidth of one to even ten GHz to transfer digital information with very low transmit power. Especially, the transmit power is smeared over the large transmit bandwidth so that UWB systems can be used without disturbing other legacy radio systems in the same frequency band. The low power is then the reason why UWB is typically a short range communication system. The similarity to the start stop method is that it also uses a single pulse with a specific time index for data transmission, which could be interpreted as the stop bit in our approach. 
Nonetheless, for the start stop bit method, 
the counter can flexible run over multiple dimensions like subcarriers, OFDM symbols in time, time shifts, frequency shifts, modulation schemes, etc. That way, we avoid the need for a very large RF bandwidth and simultaneously achieve high data rate efficiency if needed.  

SC-IM has in the meantime attracted quite some interest and this is partly due to similar reasons, which motivated us to think about the start stop method. 
The basic idea of SC-IM is derived from spatial multiplexing including permutations of on-off antenna elements, where the active and inactive antenna elements convey the intended information. 
For SC-IM the basic idea was to switch on and off frequency domain subcarriers of an OFDM symbol instead of antenna elements. Typically, a fixed subset of all the subcarriers is switched on and off, but for higher data rates also varying numbers of active subcarriers are supported. All possible combinations of on-off subcarriers - excluding permutations - define then the number of possible data bits per transmission. 
As one possible extension also modulation schemes per subcarrier are considered as well as some suitable combinations with the spatial domain, i.e, MIMO OFDM. As we will see in more detail further below the start stop method deviates in several aspects from SC-IM.  
First of all, it uses a predefined number of stop bits and, especially, a very low number of stop bits in the range of $10$ to $20$ out of $1024$ or more subcarriers. This maintains a very high sparsity and depending on the side conditions also a high energy efficiency. In addition, it is a prerequisite for the very effective time shifting and frequency shifting techniques extending the dimensions of the counter subspace. The main aspect is that a single stop bit might carry an extremely high information content including the related resource element of the counter (e.g., ten bits), the QAM bit mapping (e.g., eight bits), the time shift as well as the frequency shift (e.g., up to ten bits). Therefore, overall one stop bit might carry the information of up to 28 bits so that with very few bits a very good resource usage in bpcu can be achieved, making the overall transmit signal very sparse. These techniques require that adjacent subcarriers and time domain symbols of a stop bit are empty, which is not ensured by SC-IM. The remapping of estimated stop bits to bit sequences $ \mathbf{b}_{m} $ is straight forward and therefore simpler to implement compared to the remapping of combinatorial subcarrier indices with varying numbers.  As a further difference we will describe a simple coding method based on repetition coding of the stop bits, which again relies on the assumption of very few active and isolated subcarriers, which is not ensured for SC-IM.

\section{Basic Start-Stop Method and further Extensions}
\label{secIII}

\begin{figure}[tbp]
    	\centering
    	\includegraphics[width=1.0\columnwidth]{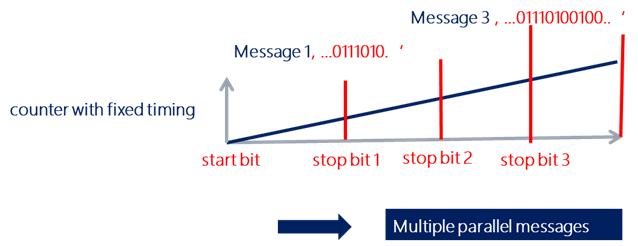}
    	\caption{Basic concept of the start stop method. }
    	\label{fig:basic_start_stop}
 \end{figure}

 Figure~\ref{fig:basic_start_stop} illustrates the basic concept of the start stop counter method. For an uplink data transmission, for example, from the UE/sensor to the gNB, the UE transmits first a start bit at a predefined time. Together with the start bit a counter is started, where the counter value is increased sequentially with a predefined clock rate. When the gNB receives the start bit, it will start its own counter using the same clock rate. Now, the UE counter is running until it reaches the specific counter value $ n_{\mathrm{c}}~\le~N_{\mathrm{c}} $ with $ N_{\mathrm{c}}~\in~\mathbb{N} $  representing the bit sequence $ \mathbf{b}~=~[ b_{1} \ldots b_{i}  \ldots   b_{N_{\mathrm{b}}}   ] $, of size $  N_{\mathrm{b}} $ transmit bits. The value of $ n_{\mathrm{c}} $ is then calculated as $ n_{\mathrm{c}}~=~\sum 2^{i-1}  b_{i} $ with  $ i~=~1 \ldots N_{\mathrm{b}} $, such that the maximum counter length is defined by the number of bits $ N_{\mathrm{b}} $ to be transmitted per bit sequence $ \mathbf{b} $. Theoretically, for extremely large counter length, the number of bits $ N_{\mathrm{b}} $ can become large as well. In that case, it would be possible to transmit a high number of bits with just two single bits, i.e, the start and the stop bit. This is the main motivation for the proposed method, i.e., to transmit with just two bits the information of $ N_{\mathrm{b}} $ bits. This results in a lossless compression ratio of $ \frac{2}{N_{\mathrm{b}}} $. For example, for a message sequence length $ N_{\mathrm{b}}~=~20 $, one would achieve a tenfold compression ratio with respect to the number of transmitted bits versus number of start-stop bits.  

In a slightly modified version, even a single stop bit will be sufficient. In this case the counter starts either at a well defined start time, or, the start of the counters is triggered by the gNB, for example, by a broadcasted trigger signal, which can start then the counters of multiple UEs. In this case the lossless compression ratio would be $ \frac{1}{N_{\mathrm{b}}} $ so that with $ N_{\mathrm{b}} $ equal to ten bits we get a factor of ten lower number of start-stop bits. 

\begin{figure}[htbp]
	\centering
	\includegraphics[width=1.0\columnwidth]{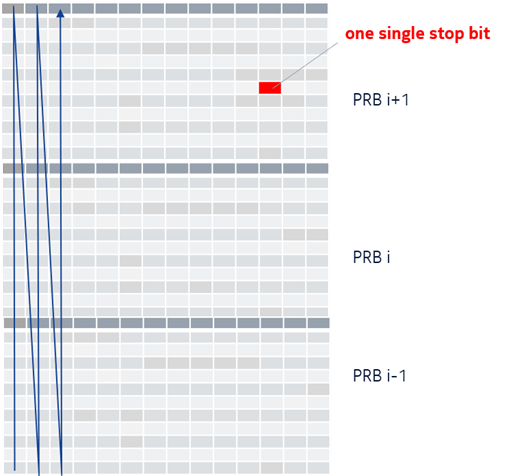}
	\caption{Allocation of counter values as frequency first for a classical 3GPP NR OFDM system over multiple PRBs of a frequency subband. }
	\label{fig:SC_symbol_grid}
\end{figure}

Here, we have assumed any possible time domain counter. However, for 3GPP NR OFDM systems, we can map the counter to the resource elements (RE) of a PRB or a set of PRBs. Figure~\ref{fig:SC_symbol_grid} illustrates the mapping for a single stop bit with a frequency first allocation of the counter values to REs. In this case, we do not need a start bit as we can define the first subcarrier in the first OFDM symbol as the start bit. 
This is beneficial since the NR OFDM system can be mostly reused as only the mapping from bits to REs has to be defined.

For practical reasons, a relatively low value for the bit sequence length $ N_{\mathrm{b}} $ is important as the size of the counter increases exponentially with $ N_{\mathrm{b}} $. For example, with $ N_{\mathrm{b}}=10 $ the counter length is $  n_{\mathrm{c,max}}(10)=2^{10}=1024 $, while for $ N_{\mathrm{b}}=20 $ it would be already $  n_{\mathrm{c,max}}(20)=2^{20}=1048576 $. A high counter length leads then to a large counting delay and, more importantly, to a low resource usage $ \gamma_{\mathrm{s}} $. 
Resource usage, or, data rate efficiency means that during the, e.g., 1024 time steps only ten bits are transmitted, while a conventional BPSK TDMA channel with the same bandwidth (i.e., same maximum counter clock length = number or REs) could transmit in the same time 1024 bits. For some sensor devices with seldom low to moderate data packets size, a low resource usage might be fine, while for more general applications like RedCap UEs \cite{REDcap} a reasonable resource usage will be important. 

\begin{figure}[tbp]
	\centering
	\includegraphics[width=1.0\columnwidth]{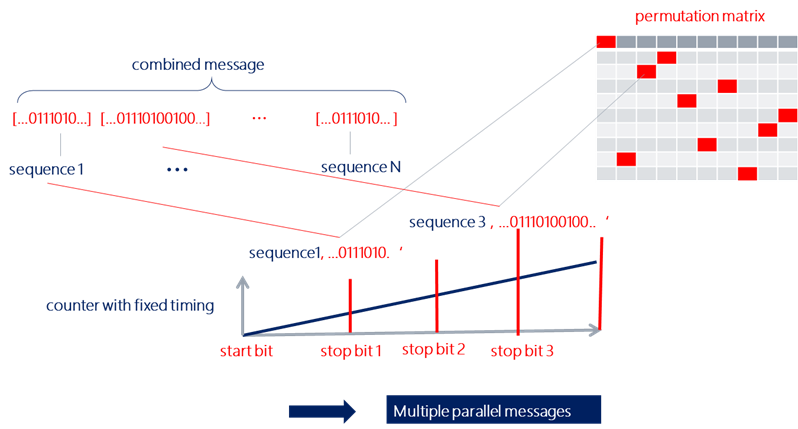}
	\caption{Illustration of start stop method transmitting multiple bit sequences in parallel.}
	\label{fig:diagram_multi}
\end{figure}

To improve the resource usage for the basic start stop method, we propose \textcolor{bvb}{to first transmit} %as a first means a transmission of 
multiple parallel bit sequences, as already illustrated in Figure \ref{fig:basic_start_stop}. The main idea is to use the same counter and to report multiple stop bits for $ M $ bit sequences  $ \mathbf{b}_{m}~=~[ b_{1} \ldots b_{i}  \ldots   b_{N_{\mathrm{b}}}   ] $ of length $ N_{\mathrm{b}} $ each so that we can report overall a $ M $ times $ N_{\mathrm{b}} $ bit sequence $ \mathbf{b}~=~[ \mathbf{b}^{T}_{1} \ldots \mathbf{b}^{T}_{m}  \ldots   \mathbf{b}^{T}_{M}   ]^{T} $. The remaining challenge is to define which stop bit belongs to which $ \mathbf{b}_{m} $ bit sequence.
The most promising solution for this purpose is to report an additional permutation matrix $ \mathbf{P} \in \mathbb{R}^{M \times M} $,
as this matrix will be sparse with only one non zero ‘1’-element per row. As illustrated in Figure~\ref{fig:diagram_multi}, the position of the ‘1’-element in a row maps then a stop bit to the index $ m $ for the related bit sequence $ \mathbf{b}_{m} $.   As a result, we can transmit $ M $ times $ N_{\mathrm{b}} $ bits with $ N_{\mathrm{b}} $ stop bits plus $ M $ reordering bits. For the case of $ M=~N_{\mathrm{b}}~=~10 $, we transmit $100$ bits with only 20 non-zero elements so that the compression ration is a factor of five. The required number of resource elements is $  N_{\mathrm{RE}}~=~2^{N_{\mathrm{b}}} + M \times M  $~=~1024~+~100~=~1124. Then, the related $ \gamma_{\mathrm{s}}=\frac{100}{1124}=0.089 $ is now improved compared to the transmission of a single bit sequence with $ M $~=~1 and  $ \gamma_{\mathrm{s}}=\frac{10}{1024}=0.01 $, but still below ten percent and therefore quite low. 

\section{Enhanced Start-Stop Method }
\label{secIV}

The basic start stop method provides a high energy efficiency, but achieves only a very limited user data rate compared to the number of data resource elements, i.e., its data rate efficiency $ \gamma_{\mathrm{s}} $ in bpcu is low. Therefore, in a first enhancement, we propose the concept of fractional time and frequency shifts for the stop bits. A second enhancement proposes an enhanced coverage mode using a stop bit repetition mode.

\subsection{Enhanced Data Rate Efficiency}
\label{subsecIIIa}

\begin{figure}[htbp]
	\centering
	\includegraphics[width=1.0\columnwidth]{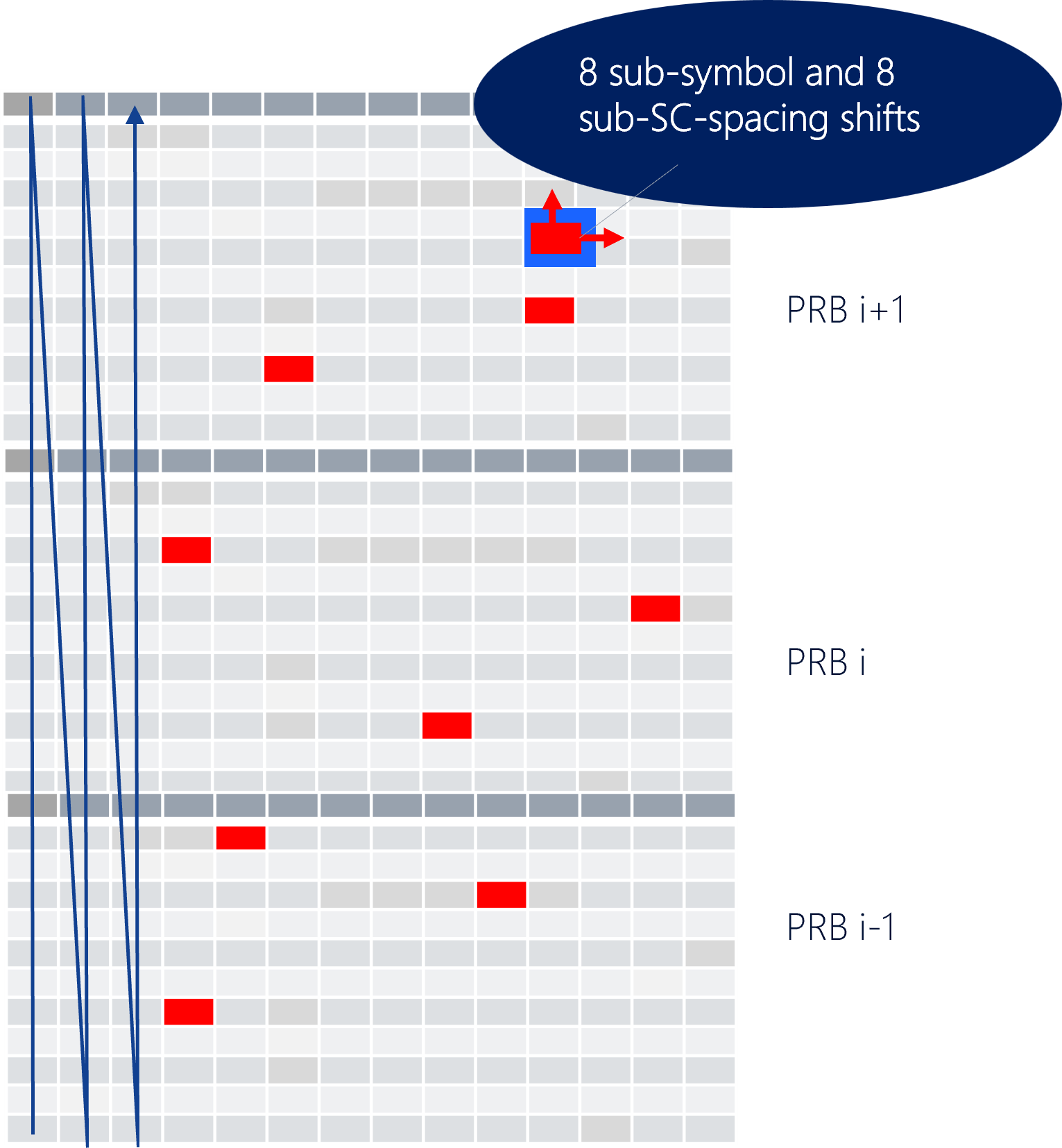}
	\caption{Fractional time and frequency shifts of stop bits. }
	\label{fig:TS_FS}
\end{figure}

As discussed in Section~\ref{secII}, the resource efficiency $ \gamma_{\mathrm{s}} $ of a multi sequence start stop communication link is about a factor of ten below that of a conventional NR system. This limits - similar as known for SC-IM systems - the usability for high end applications like high capacity dMIMO systems. For that reason, we add a further dimension for data transmission, i.e., the fractional time shift and frequency shift (TS-FS) domain, which is illustrated in Figure~\ref{fig:TS_FS}. For the fractional time shift, the RE carrying the stop bit, the related subcarrier $ i $ of OFDM symbol $ l $, shifts this signal by a fraction $ n_{\mathrm{t}}   \delta t $ of the OFDM symbol length $ t_{\mathrm{OFDM}} $ to the left, where $ \delta t $ might be, e.g., 1/16 from $ t_{\mathrm{OFDM}} $. 

\begin{figure}[tbp]
	\centering
	\includegraphics[width=1.0\columnwidth]{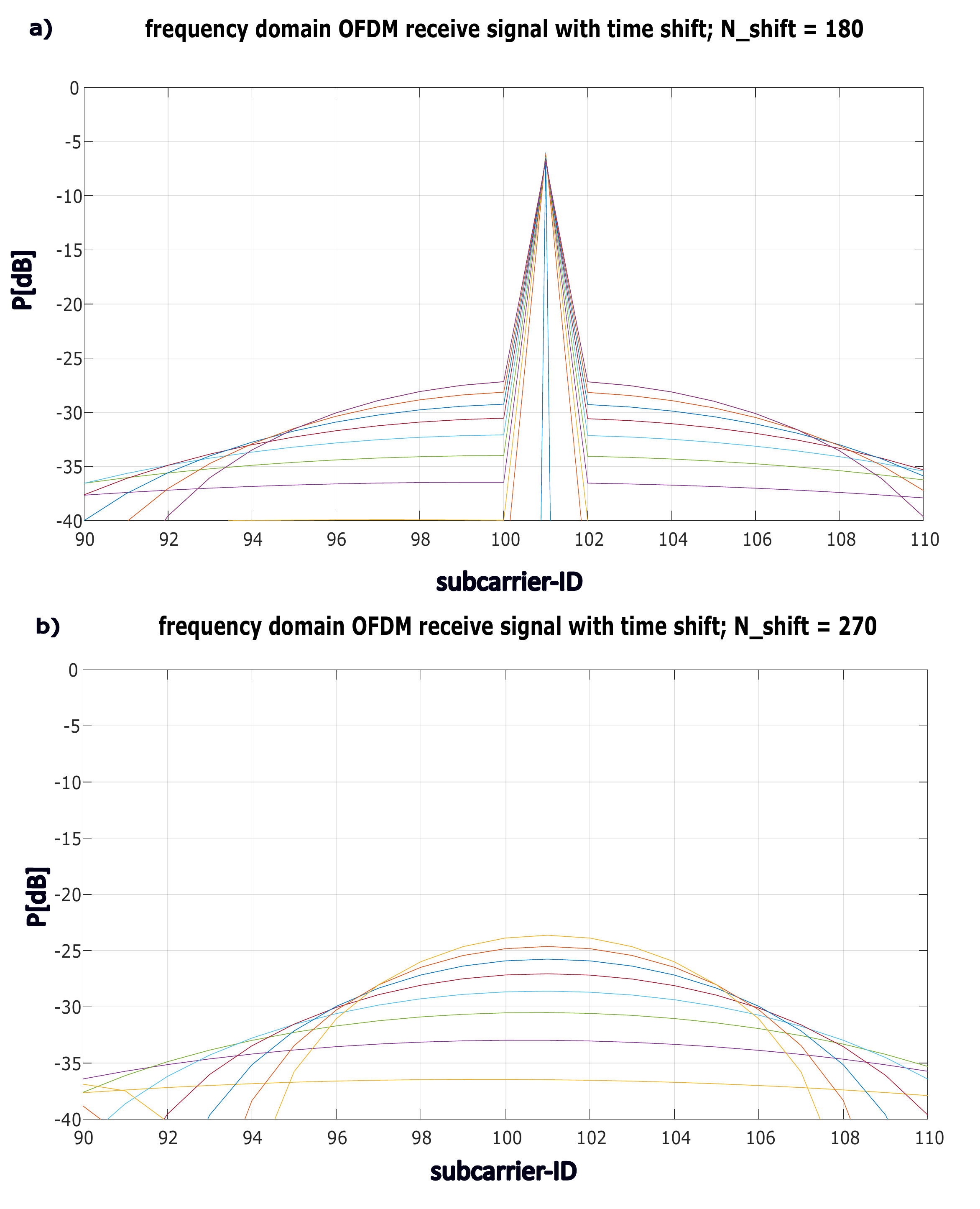}
	\caption{Inter symbol interference for 16 different fractional time shifts and $\delta t$ by shifted  a) 180 b) 270 samples}
	\label{fig:TD}
\end{figure}

Note that such a fractional time and frequency shifting leads to inter symbol and inter carrier interference.  Figure~\ref{fig:TD} illustrates the effect of a fractional time shift on the stop bit itself as well on the adjacent following OFDM symbol for a time shift of $ \delta t  = 270$ out of $2048$ samples for the full OFDM symbol.

\begin{figure}[htbp]
	\centering
\includegraphics[width=1.0\columnwidth]{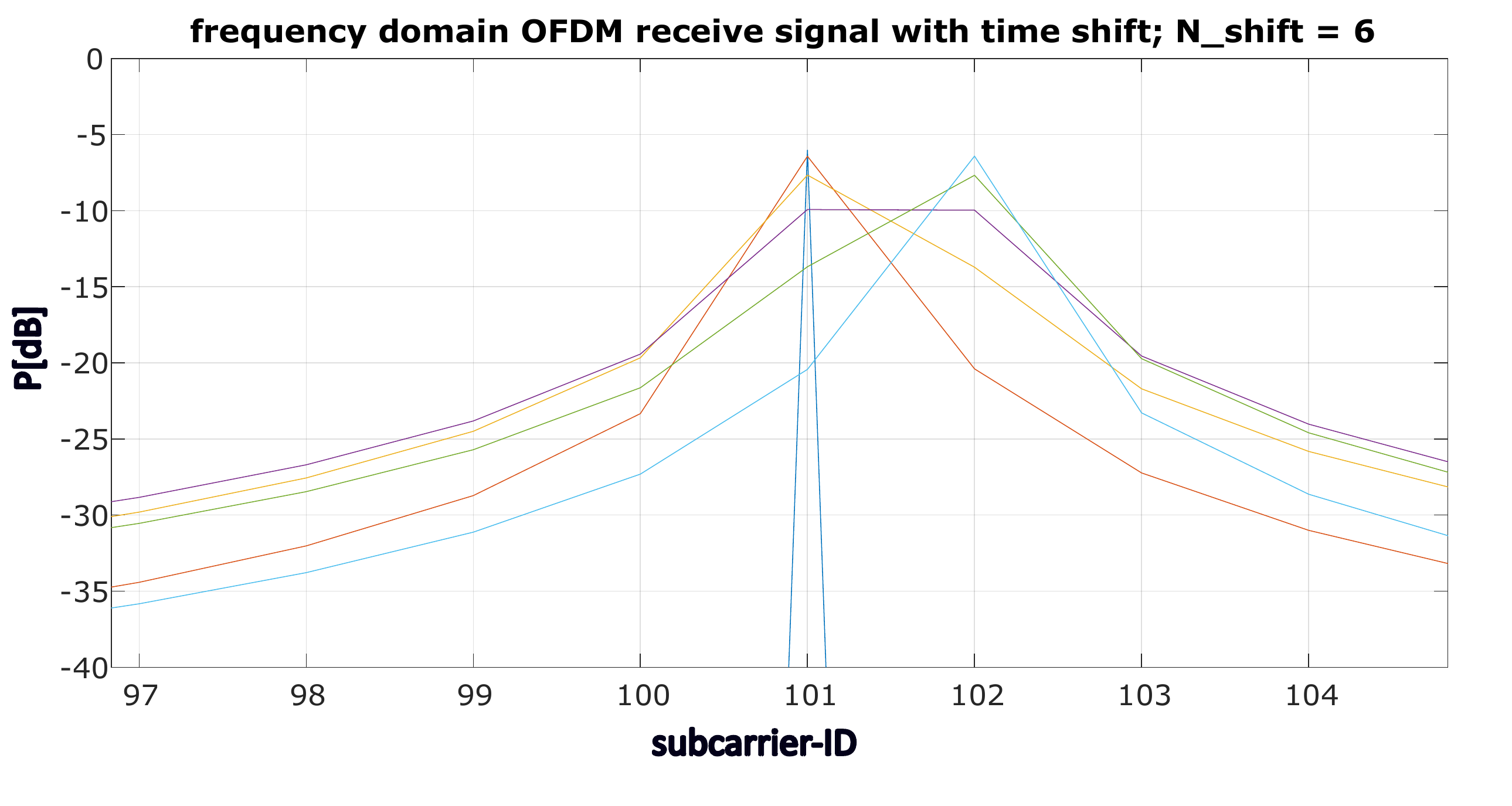}
	\caption{Inter symbol interference for 6 different fractional frequency shifts $\delta f $. Note that the stop bit subcarrier without any frequency shift is $ i $ = 101 in the center.}
	\label{fig:frequency_shift}
\end{figure}

Similarly, Figure~\ref{fig:frequency_shift}  illustrates the effect of a fractional frequency shift $ \delta f $ of the stop bit subcarrier on the receive power at the stop bit subcarrier as well for the adjacent subcarriers. 

More research is needed to analyze the optimum set of parameters as well as the best receiver demodulation methods. Especially, it is reasonable to evaluate the time and frequency shifted stop bits over a TS-FS grid covering the RE of the stop bit plus that of the adjacent subcarriers and OFDM symbols like the blue area in Figure~\ref{fig:TS_FS}. Then a combined evaluation over this TS-FS grid leads to the best demodulation performance. AI/ML based methods seem to be well suited to learn the typical receive signal impacts for different time frequency shift combinations $ n_{\mathrm{t}} \delta t $ and $ n_{\mathrm{f}} \delta f $. As we can see in Figure~\ref{fig:TS_FS} one single stop bit affects multiple REs. Therefore, in the affected blue area no other stop bit should be received to avoid inter stop bit interference.

This can be accomplished quite easily by a modified counter, which skips the REs in the blue area, i.e., the counter continues with the first RE after the blue area. When this simple rule is predefined and known, then the receiver can adapt its counter correspondingly for the reconstruction of the bit sequences $ \mathbf{b}_{m} $. The skipping of the blue REs requires that the overall number of counter REs $ N_{\mathrm{RE}} $ has to be increased, i.e., for each of the $ M $ stop bits by $ N_{\mathrm{TS-FS}} $ bits, where $ N_{\mathrm{TS-FS}} $ defines the number of REs of the blue TS-FS grid area. A typical number for $ N_{\mathrm{TS-FS}} $ might be 4. In that case, the overall counter length would be $ N_{\mathrm{RE}}~=~2^{N_{\mathrm{b}}} + M \times M  + M N_{\mathrm{TS-FS}} $~=~1024~+~100+40~=~1164. Hence, the counter length is increased only by few percent. Such a counter length would fit, e.g., into a set of 8 PRBs a 168 REs, which have 1344 REs and leave room for 22 demodulation reference signal REs. Anyway, the best possible protocol design is for further study.  

Applying a fractional TS-FS grid on a stop bit allows to increase the data rate efficiency $ \gamma_{\mathrm{s}} $ depending on the size of the TS-FS grid. For example, to achieve the same $ \gamma_{\mathrm{s}} $ as for a conventional NR OFDM system, we need a factor of ten higher number of bits per stop bit. This requires a TS-FS grid size of $ 2^{10}~=~1024 $, i.e., a number of fractional time shifts $ N_{\mathrm{t}} $ and frequency shifts $ N_{\mathrm{f}} $  equal to 32. The TS-FS grid size is then $ 32 \times 32 $. Obviously, such an allocation is not for free and will require a high SINR to be able to accurately separate between different time and frequency shifts. As a rough estimate, we expect a $ \Delta $SINR per bit of about 3dB, i.e., for ten bits the SINR has to be increased by $ \Delta $SINR~=~30dB. 

This can be partly compensated by the possible power boost $ P_{\mathrm{boost}} $ of stop bits compared to the conventional fully loaded OFDM system with $1024$ REs as we have only $ 2 M~=~20 $ active stop bits. This leads to a possible power boost of $ P_{\mathrm{boost}}~=~10 \log(1024/20)~=~17$~dB, which leaves a gap of 30dB-17dB~=~13dB. Such a quite large gap of $13$~dB might be acceptable in some cases or specific scenarios with extreme high SINR conditions like short distance industry applications. Another approach is to reconfigure the start stop bit parameters, i.e., to increase the number of parallel transmitted messages $ M $ from 10 to 20. Then, the TS-FS grid has to support only 5 bits, i.e., the TS-FS grid size is reduced to $ 2^{5}~=~32 $ which can be almost achieved by $ N_{\mathrm{t}}~=~5 $ and  $ N_{\mathrm{f}}~=~6 $. The expected  $ \Delta $SINR is then  $ 5*3~=~15 $dB, which can be almost compensated by the slightly lower power boost of $ P_{\mathrm{boost,20}}~=~10 \log(1024/40)~=~14 $dB. Finding the overall best fitting parameter setting is for future research and depends also on other side conditions    

\subsection{Enhanced Coverage}
\label{subsecIIIb}

Enhanced coverage is not in the focus of this paper, but the main idea is that the start stop bit methods might have some benefits. For conventional deep coverage UEs it is common to apply a strong repetition coding \cite{Rcoding} to overcome the large pathloss, e.g. of UEs located in a cellar. Assuming we have a message of $1024$ bits, then these $1024$ bits have to be repeated, for example $20$ times, leading to a very inefficient resource usage $ \gamma_{\mathrm{s}} $. 
The start stop methods might provide here some unique benefits as we might exploit the power boost $ P_{\mathrm{boost}} $ of, e.g., $14$ to $17$dB as explained in Section~\ref*{subsecIIIa}. 
In addition, we may consider a very efficient repetition coding, where only the stop bits are repeated multiple times. 
This can be very resource effective, as even with $ R~=~20 $ stop bit repetitions the overall length of the counter will just increase to $ N_{\mathrm{RE}}~=~2^{N_{\mathrm{b}}} + M \times M  + M R $~=~1024~+~100~+~200~=~1324. Note that such a concept has some implications requiring a more careful analysis, but this is beyond the scope of this paper.

\begin{figure}[tbp]
	\centering
	\includegraphics[width=1.0\columnwidth]{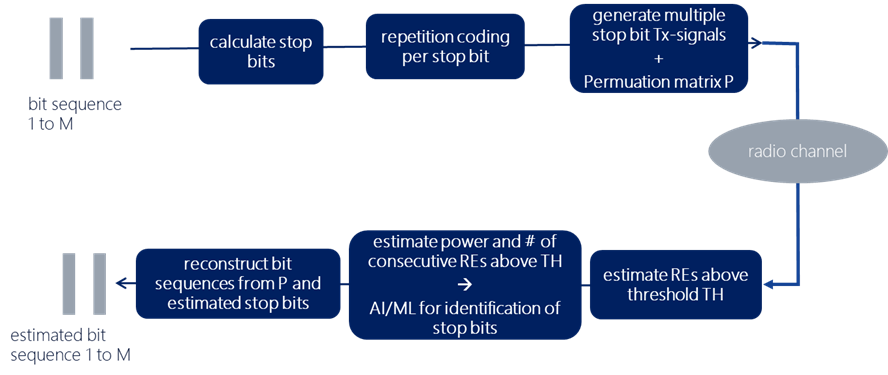}
	\caption{Basic block diagram of a start stop bit system including multiple messages $ M $ and repetition coding of stop bits. }
	\label{fig:multi_block}
\end{figure}

Figure~\ref{fig:multi_block} provides a block diagram with the main steps of a start stop system at the transmit and the receive side including the transmission of multiple messages as well as the repetition coding of start stop bits. At the receiver we assume already an AI/ML based demodulator.

\section{Sparse dMIMO System Concept}
\label{secV}

Let us shortly discuss the potential application of the \emph{start stop} bit method to Joint Transmission Cooperative Multipoint Transmission (JT CoMP) systems leading to a so called sparse dMIMO concept. JT CoMP has a long history of research \cite{Role}, but the practical impact has been small so far. Reasons for that are the high complexity as well as moderate to marginal system level gains for realistic as well as ideal channel estimation assumptions. The goal of sparse dMIMO is to overcome some of the main issues, i.e., i) the significant inter cooperation interference, which often counteracts most of the ideal performance gains and ii) the high precoder complexity for large channel matrices.

\begin{figure}[tbp]
	\centering
	\includegraphics[width=1.0\columnwidth]{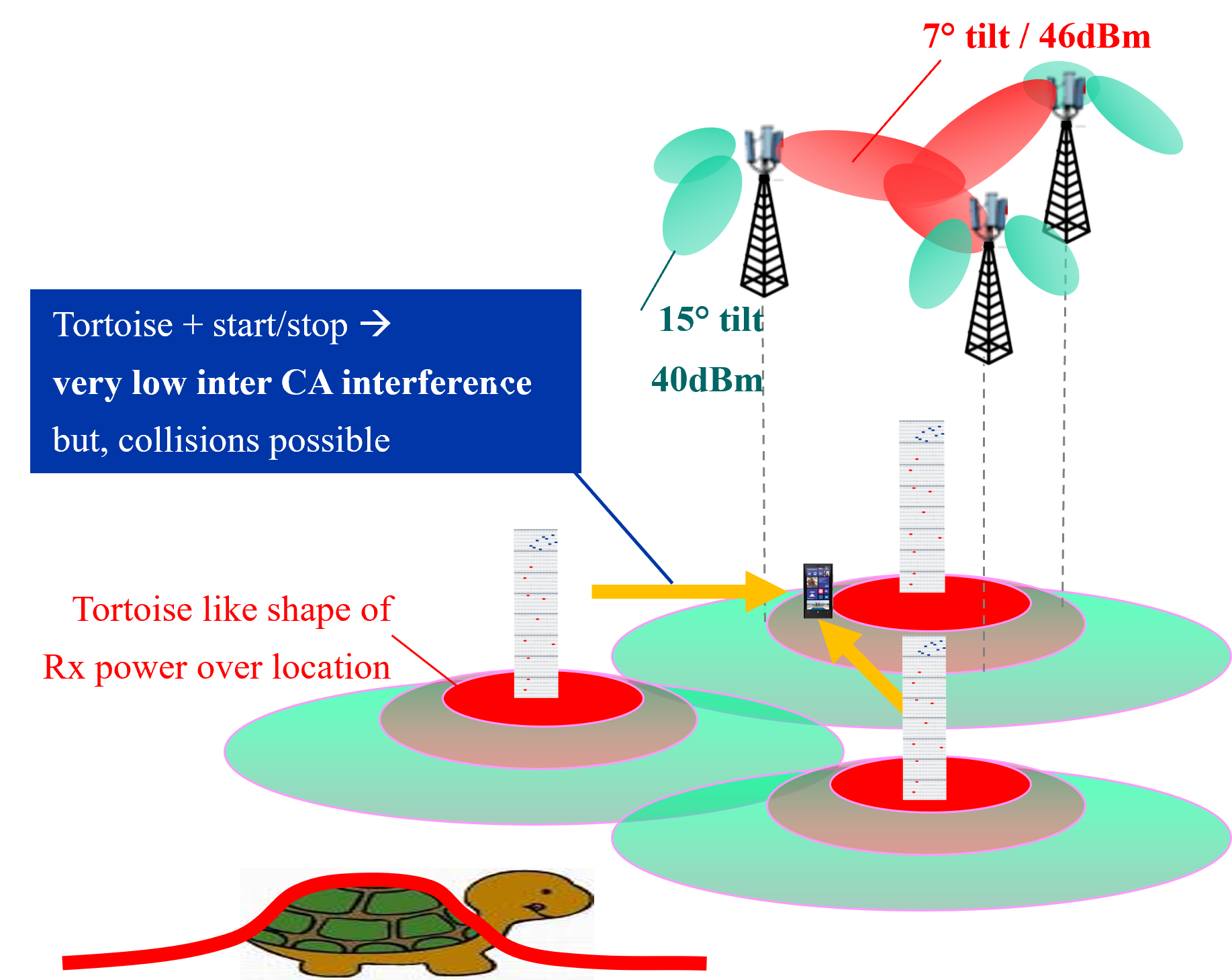}
	\caption{Sparse dMIMO scenario with low inter cooperation area interference due to less than two percent of active resource elements per data link. }
	\label{fig:dMIMO_scenario}
\end{figure}

Figure~\ref{fig:dMIMO_scenario} illustrates the basic dMIMO system, which is basically the combination of a JT CoMP system with the start stop bit method. We assume conventional cooperation areas of nine cells applying the tortoise method to achieve already a quite good inter cooperation decoupling as explained in \cite{tortoise}. This is a UE unaware wideband method, which reduces the general inter cooperation area interference floor by different tilting for in and outbound beams. At top of this, we assume the start stop method and exploit the extreme sparsity of active resource elements. For example, for $ N_{\mathrm{b}}~=~10 $ and $ M~=~10 $ we have  $ N_{\mathrm{RE}}~=~2^{N_{\mathrm{b}}} + M \times M  + M N_{\mathrm{TS-FS}} $~=~1024~+~100~+~40~=~1164 REs, from which \textcolor{bvb}{only} $ 2 M~=~20 $ REs are active. That leads to a very high sparsity level of $ 20/1164~=~0.017 $. Obviously, only active bits will generate inter cooperation \textcolor{bvb}{area} interference. 
This might reduce the inter cooperation area interference for sparse dMIMO significantly by a potential factor of about 50. But, we should note that the statistics of the inter cooperation area interference is now significantly different to that of conventional JT CoMP systems. In addition, we have quite a different decoding process at the receiver side, which might require novel interference mitigation methods. Optimum usage of the high sparsity is for further research and a first step will be to analyze the typical inter cooperation area statistics. 

Besides the reduced inter cooperation area interference, sparse dMIMO has also a severe impact to the overall precoder processing. Here, we can benefit from the sparse number of active subcarriers and OFDM symbols per PRB. We expect two benefits, i) a more efficient calculation of the precoder matrices like the Moore Penrose Pseudo inversion of the channel matrix H $ \mathbf{W}_{i}~=~\mathrm{pinv}(\mathbf{H}_{i}) $, and more importantly ii) a reduced effort for the matrix multiplications  $ \mathbf{W}_{i} \: \mathbf{\tilde{x}}_{j} $. Note that, the lower processing effort is due to the sparsity of the transmit vector $ \mathbf{\tilde{x}}_{j} $, where the tilde indicates the mapping of the original bit sequences $ \mathbf{b} $ for the user data vector $ \mathbf{x}_{j} $ to the corresponding start stop bit allocation in the vector $ \mathbf{\tilde{x}}_{j} $. Many entries of $ \mathbf{\tilde{x}}_{j} $ will be for many subcarriers $ i $ zero, so that the related matrix multiplication can be skipped. Similar as for the inter cooperation area interference, one can expect potential factor of about 50 lower processing complexity, but the details will have to be checked more carefully in the future.

\section{Conclusion}
\label{sec:conclusion}

We have proposed the start stop bit method, which promises energy efficient data transmission for the next generations of mobile wireless systems. 
We proposed the start stop bit methods as well as several extensions to the basic scheme which can achieve energy efficiency, and high data rate efficiency in bpcu, which affects the user throughput and/or the capacity in b/s/Hz/cell.

A specific coding scheme has been discussed, which promises as further feature very high coverage with low resource usage. 

In the future, interesting will be more research related to sparse dMIMO system level aspects  like beneficial inter cell interference conditions due to sparse signal transmission, or, the lowest possible precoder signal processing complexity.  
Similarly, implementation aspects like channel estimation, synchronization, multi user scheduling, and the possible standardization impact, needs more consideration.

\bibliographystyle{IEEEtran}
\bibliography{REF_WRC2024}
 
\end{document}